\documentclass[aps,pra,nofootinbib,preprint,amsmath,amssymb,floatfix]{revtex4}
\usepackage{color}
\usepackage{graphicx}

\begin{document}
\newcommand{\tc}{\textcolor}
\title{Critical properties of the hierarchical reference theory}         

\author{Johan S. H\o ye$^1$ and  Enrique Lomba$^2$}        
\affiliation{$^1$Institutt for Fysikk,  NTNU, N-7491 Trondheim, Norway\\ 
$^2$Instituto de Qu{\' \i}mica F{\' \i}sica Rocasolano, CSIC,
  Calle Serrano 119, E-28026 Madrid, Spain}

\date{\today}          

\begin{abstract}

The critical region of the hierarchical reference theory (HRT) is
investigated further. This extends an earlier work by us where the
critical properties of the HRT were concluded indirectly via another
accurate but somewhat different theory, the self-consistent
Ornstein-Zernike approximation (SCOZA), and numerical work. In the
present work we perform our analysis directly upon the HRT partial
differential equation to establish ordinary differential equations for
the subleading scaling contributions. Again we find that the HRT
critical indices in three dimensions are simple rational numbers.

\end{abstract}
\maketitle

\bigskip
\section{Introduction} 
\label{secintro}

It is a pleasure for us to contribute this article to the festschrift devoted to the memory of Professor Lesser Blum. We both have collaborated with him on problems in the statistical mechanics of fluids. Especially the study of electrolytes and solutions of the Ornstein- Zernike equation were central in this respect.

The present contribution is a further study of the critical properties of fluids. Thus we extend our previous analysis of the HRT equation \cite{hoye11}. The HRT was introduced as a new and accurate method to evaluate the equation of state of fluids \cite{parola85}. It was inspired by the momentum-space renormalization group theory that was developed by Wilson and Kogut \cite{wilson74}. This method came from field theory. So with HRT one for a given temperature adds Fourier components or wave vectors to the Fourier transform of the perturbing interaction until the full interaction is obtained. Further the pair correlation function is assumed to be of the Ornstein-Zernike form \cite{ornstein14} with a free parameter to be determined from self-consistency between corresponding expressions for the free energy and compressibility. This can be expressed with a partial differential equation.

From numerical work on HRT accurate results for the equation of state of continuum fluids came out. Especially, the critical region was described well, and very good results for the critical indices were found \cite{parola85,parola93,parola95,ionescu07,parola08,parola09}.

Another accurate approach to obtain the equation of state was the Ornstein-Zernike approximation (SCOZA) developed by H{\o}ye and Stell \cite{hoye77,hoye84,hoye85}. Again the Ornstein-Zernike equation of fluid theory with its mean spherical approximation (MSA) was a basis. With MSA the direct correlation function outside hard cores is assumed to the perturbing potential times $-\beta$ with $\beta=1/(k_B T)$ where $k_B$ is Boltzmann's constant and $T$ is temperature.

With SCOZA the $\beta$ is replaced by an effective inverse temperature as a free parameter to be determined via thermodynamic self-consistency between the internal energy and compressibility. Again very accurate results came out, also in the critical region, and various generalizations have been dealt with \cite{dickman96,borge98,pini98,pini98a,pini98b,hoye00,grollau01,scholl03a,scholl03b,scholl05,betancourt08}.

A unification of HRT and SCOZA was made by H{\o}ye and Reiner \cite{reiner05,hoye07}. The unified problem was further analyzed in the critical region \cite{hoye09}. This latter problem with two free parameters was found to be essentially a sum of the HRT and SCOZA problems. Since these two theories qualitatively have somewhat different critical behavior, a problem was how these different behaviors  could be reconciled, if possible. It was found that the HRT part would dominate in the critical region. But, according to the analysis performed, reconciliation with SCOZA implied presence of subleading terms scaling terms. This was investigated more closely by analytic and accurate numerical work in Ref.~\cite{hoye11} by solution of the HRT partial differential equation. It was found that the HRT critical indices were simple rational numbers.

In the present work we will investigate the critical properties of HRT directly from the HRT equation without indirect arguments via the SCOZA equation. To do so the solution of the partial differential equation will be split into leading and subleading contributions in the critical region. These contributions will fulfill ordinary differential equations.

In Section \ref{sec1} we will write down the simplified form of the HRT equation in the critical region and reconsider shortly the special case where it coincides with the SCOZA.

In Section \ref{sec2} the HRT equation  is expanded in subleading contributions, and we specialize to a simple situation with respect to the cutoff function $L$.
 
Then in Section \ref{sec3} the fixed point solution, which is the leading scaling function, is established as a simple analytic expression.

In Section \ref{sec4} subleading contributions to the critical isotherm are analyzed while in Section \ref{sec5} the leading and subleading contributions for deviation from the critical temperature are considered.

In Section \ref{sec6} numerical results for the subleading contributions are obtained, and a summary is made in Section \ref{sec7}.

\section{HRT equation} 
\label{sec1}

In our earlier work we found the critical properties of the HRT for fluids via somewhat indirect analytic arguments and accurate numerical analysis of the HRT equation \cite{hoye11}. The indirect arguments concerned the use of the related equation for the self-consistent Ornstein-Zernike approximation (SCOZA). By adjusting the internal energy by a parameter $\nu$ these two equations should give the same result. Then Eq.~(23) of Ref.~\cite{hoye11} for $\nu$ was derived. From this differential equation it was concluded that $\nu$ would contain only scaling terms already present in the HRT and/or SCOZA theories. This had the implication that the HRT critical indices were simple rational numbers with given index $\delta=5$. This conclusion was supported by numerical solution of the HRT equation with high accuracy.

In view of these results it should in some way or other be possible to show the resulting critical behavior directly from the HRT partial differential equation itself, provided the previous analysis was correctly performed. The equation is non-linear and corresponds to the description of a non-linear diffusion of heat conduction process. Since it is non-linear, it is not obvious how to analyze it. But we are led by what to expect in view of earlier analysis and results obtained. Thus we aim at expansion of the solution of the equation in its leading and subleading scaling functions where the latter will be solutions of linear ordinary differential equations.

The HRT equation, Eq.~(33) in Ref.~\cite{hoye11}, is
\begin{equation}
y_Q+\frac{\partial}{\partial m}(Ly_m)=0.
\label{1}
\end{equation}
Here $y$ is the inverse susceptibility or compressibility, $m$ is magnetization or order parameter, and $Q$ is the wave vector cutoff of the perturbing interaction in Fourier space. Further $L$ with scaling form $L=L(y/Q^2)$ is the cutoff function. Eq.~(\ref{1}) is the simplified form of the HRT equation in the critical region. Away from the critical point one should have mean field behavior; so for some finite $Q$ one has from Eq.~(5) of Ref.~\cite{hoye11} the boundary condition  
\begin{equation}
y=a'm^2+b'
\label{2}
\end{equation}
with coefficients $a'$ and $b'$ where $b'$ will depend upon $Q$ and temperature deviation from the critical one. In Eq.~(\ref{1}) the subscripts $Q$ and $m$ indicate first derivatives with respect to these variables.

As in Ref.~\cite{hoye11} it is convenient to exchange the roles of the variables $m$ and $y$. So with $u=m^2$ Eq.~(\ref{1}) turns into Eq.~(27) of the reference
\begin{equation}
u_Q-2L_e=0, \quad L_e=2\frac{u}{u_y}L_y- 2\frac{u u_{yy}}{u_y^2}L+L
\label{3}
\end{equation}
where one and double subscripts $y$ mean first and second derivatives with respect to it.

As a sidestep here, if (\ref{1}) is changed into the HRT equation of the spherical model, it will be $y_Q+(1/m)L y_m=0$ with corresponding transformed equation $u_Q-2L=0$. This equation solves the spherical model exactly and has solution $u=-2J+f(y)$ with $L=-\partial J/\partial Q$ \cite{hoye07,hoye09}. With Eqs.~(\ref{1}) and (\ref{3}) the situation is much more demanding.

However, with $L$ given by Eq.~(26) of Ref.~\cite{hoye11} the situation is again simple due to equivalence to the corresponding SCOZA problem. In this case
\begin{equation}
L=\frac{Q}{\sqrt{y+Q^2}}.
\label{4}
\end{equation}
Then one finds the fixed point solution 
\begin{equation}
u=u_0=2\sqrt{y+Q^2}.
\label{5}
\end{equation}
The subleading contribution to this is found by putting
\begin{equation}
u=u_0+u_1+\cdots
\label{6}
\end{equation}
($u_0\sim Q$, $u_1\sim Q^2$) for small $Q$ to obtain Eq.~(30) of Ref.~\cite{hoye11}
\begin{equation}
u_{1Q}+4Q(u_{1y}+2(\sqrt{y+Q^2})u_{1yy})=0.
\label{7}
\end{equation}
Its solution, consistent with condition (\ref{2}), is
\begin{equation}
u_1=2c(y+3t-2Q^2)
\label{8}
\end{equation}
where $c$ is a constant while $t$ ($<0$) is deviation from inverse critical temperature. (The coefficient of the $t$ term was determined from the corresponding SCOZA equation with $\nu=0$.)

\section{Expansion in subleading contributions} 
\label{sec2}

Now we will consider the more general situation where two levels of subleading contributions of importance is present. Then the transformed  Eq.~(\ref{3}) is to be expanded to second order. To simplify we here will consider the specific case with
\begin{equation}
L=L(z)=\left\{
\begin{array}{ll}
1, &z<1,\\
0, &z>1,
\end{array} 
\right.
\label{110}
\end{equation}
where
\begin{equation}
z=\frac{y}{Q^2}.
\label{111}
\end{equation}
With this the $L_y$ term vanishes except at $z=1$ where this will be taken care of by a boundary condition. So we find ($u=u_0+u_1+\cdots$)
\begin{eqnarray}
\label{112}
\frac{u u_{yy}}{u_y^2}&=&\frac{u_0 u_{0yy}}{u_{0y^2}}(1+c+d),\\
\label{113}
c&=&\frac{u_1}{u_0}+\frac{u_{1yy}}{u_{0yy}}-2\frac{u_{1y}}{u_{0y}},\\
\label{114}
d&=&\frac{u_1}{u_0}\frac{u_{1yy}}{u_{0yy}}+3\left(\frac{u_{1y}}{u_{0y}}\right)^2-2\frac{u_1}{u_0}\frac{u_{1y}}{u_{0y}}-2\frac{u_{1y}}{u_{0y}}\frac{u_{1yy}}{u_{0yy}}.
\end{eqnarray}
Since the quantity $c$ can be expanded to contain both first and second order contributions, it can be split in two parts such that 
\begin{equation}
u_1\rightarrow u_1+u_2
\label{115}
\end{equation}
while $d$ of second order may be kept unchanged.

Now we will focus upon the critical isotherm with temperature parameter $t=0$. Then the functions $u_i$ ($i=0,1,2$) are assumed to have the scaling forms
\begin{eqnarray}
\nonumber
u_0&=&Q f_0(z)\\
u_1&=& Q^{3/2} f_1 (z)
\label{116}\\
\nonumber
u_2&=&Q^2 f_2(z).
\end{eqnarray}
The form of the fixed point solution $u_0$ is obvious; the $u_2$ should be able to satisfy the mean field boundary condition (\ref{2}) with needed  assistance from $u_1$ by which the latter should produce the bilinear term $d$ to go along with $u_2$. This determines the exponent 3/2 of the prefactor of the scaling form of $u_1$.

With Eq.~(\ref{116}) inserted in the HRT Eq.~(\ref{3}) we obtain the equations
\begin{equation}
\frac{1}{2}f_0-zf_0^\prime +2F_0-1=0
\label{117}
\end{equation}
\begin{equation}
\frac{3}{4}f_1-zf_1^\prime +2F_0\left(\frac{f_1}{f_0}+\frac{f_1^{\prime\prime}}{f_0^{\prime\prime}}-2\frac{f_1^\prime}{f_0^\prime}\right)=0
\label{118}
\end{equation}
\begin{equation}
f_2-zf_2^\prime +2F_0\left(\frac{f_2}{f_0}+\frac{f_2^{\prime\prime}}{f_0^{\prime\prime}}-2\frac{f_2^\prime}{f_0^\prime}+\frac{d}{Q}\right)=0
\label{119}
\end{equation}
\begin{equation}
F_0=\frac{f_0 f_0^{\prime\prime}}{f_0^{\prime\, 2}}
\label{120}
\end{equation}
for $z<1$ where primes mean derivatives with respect to $z$. For $z>1$ the equations are simply
\begin{equation}
u_{iQ}=0 \quad (i=0,1,2).
\label{121}
\end{equation}

\section{Fixed point solution} 
\label{sec3}

For the case considered the exact fixed point solution can be found. This is not obvious from Eq.~(\ref{117}). But with $L=1$ the equation is the same as the linear heat equation for $y<Q^2$ while outside this region there is no conduction ($L=0$). Thus, despite the moving boundary $y=Q^2$, the scaling solution, when assumed to be a power series, can be easily found as (for $y<Q^2$)
\begin{equation}
y=-\frac{1}{4}Q^2+\frac{1}{2}Qm^2-\frac{1}{24}m^4
\label{122}
\end{equation}
while for $y>Q^2$ the $y=(1/36)m^4$. The boundary between the two regions is located at $m^2=6Q$ where heat conduction stops such that 
\begin{equation}
\frac{\partial y}{\partial m}=0.
\label{123}
\end{equation}
[It can be seen that the integral of $y$ for $y<Q^2$ is the same as the one of $m^4/36$, i.e.~heat energy is conserved when $Q$ decreases.]
From this follows
\begin{equation}
u_0=Qf_0=m^2=Q(6-\sqrt{24(1-z)}\,),
\label{124}
\end{equation}
\begin{equation}
f_0^\prime=\frac{12}{\sqrt{24(1-z)}}, \quad f_0^{\prime\prime}=\frac{144}{(24(1-z))^{3/2}},
\label{125}
\end{equation}
\begin{equation}
F_0=\frac{1}{\sqrt{24(1-z)}}f_0.
\label{126}
\end{equation}
This is to be inserted in Eqs.~(\ref{118}) and (\ref{119}) to obtain explicit equations for $f_1$ and $f_2$. The quantity $d$ as given by Eq.~(\ref{114}) will depend upon the solution for $f_1$.

It can be noted that the scaling of the fixed point solution determines the critical index for the critical isotherm $y\sim m^4 \sim m^{\delta-1}$, i.e.~$\delta=5$. In this respect the subleading contributions are corrections to the critical isotherm towards mean field behavior away from the critical point. But indirectly they also influence the temperature dependence of critical behavior and thus the remaining critical indices as we will find.

\section{Analysis of subleading contributions} 
\label{sec4}

The equations for $f_1$ and $f_2$ are linear second order equations. Thus they each have two independent solutions. One condition is for $Q\rightarrow 0$ or $z=\frac{y}{Q^2}\rightarrow\infty$. With Eq.~(\ref{121}) the obvious solutions for $z>1$ are
\begin{equation}
f_0=6\sqrt{z}, \quad f_1\propto  z^{3/4}, \quad f_2\propto z.
\label{128}
\end{equation}
In the general situation $L$ will be non-zero, but decaying for $z>1$. Then a second solution will be present too. It will diverge rapidly for increasing $z$. In the limit $L\rightarrow 0$ the speed of its divergence will be infinite. This divergent part is discarded by boundary condition (\ref{123}) at $z=1$. With this only one combination of the two solutions of each differential equation is left. It should match the given boundary condition (\ref{2}) (or something similar) around $u\approx u_0=0$. In the present case with solution (\ref{124}) this is for
\begin{equation}
z=\frac{y}{Q^2}=-\frac{1}{2}.
\label{129}
\end{equation}
But the remaining solution of $f_2$ alone can not in general fulfill a given boundary condition. Thus the intermediate solution $f_1$ is generated by which a combination of the two solutions will be satisfactory for a given initial value of $Q$, except for possible higher order terms that are disregarded.

Near $z=-1/2$ the two solutions can be found explicitly. With $z=w-1/2$ one for small $w$ finds
\begin{equation}
f_0=2w, \quad f_0^\prime=2, \quad f_0^{\prime\prime}=\frac{2}{3}, \quad F_0=\frac{1}{3}w.
\label{130}
\end{equation}
Keeping only the leading parts of coefficients for simplicity, Eq.~(\ref{118}) becomes
\begin{equation}
\left(\frac{3}{4}+\frac{2}{3}\right) f_1+\frac{1}{2}f_1^\prime+wf_1^{\prime\prime}=0.
\label{131}
\end{equation}
This equation has solutions $1+aw+\cdots$ and $\sqrt{w}+\cdots$ (with $a$ constant), i.e.
\begin{equation}
f_1=A_1(1+a_1w)+B_1\sqrt{w}
\label{132}
\end{equation}
where $A_1$ and $B_1$ are constants. The equation for $f_2$ will be similar with 3/4 replaced by 1. With this its solution will be of the same form
\begin{equation}
f_2=A_2(1+a_2w)+B_2\sqrt{w}.
\label{133}
\end{equation}
However, the equation for $f_2$ also contains a $d$ term. As long as it is finite it should not influence form (\ref{133}), only the coefficients $A_2$ and $B_2$. But with expression (\ref{132}) the $d$ will diverge as $w\rightarrow0$. This violates our expansion of the HRT equation in this limit as the $d$ also will have some influence upon solution (\ref{133}). Thus we will assume $f_1$ small such that $d$ (and $d_t$ below) mainly can be disregarded.

For the problem of interest the ratio $B_i/A_i$ ($i=1,2$) is determined by the condition at $z=1$ (or in general by the large $z$ behavior (\ref{128})). Thus with $B_2\neq0$ boundary condition (\ref{2}) can not be fulfilled with $f_2$ alone. But it can be combined with $f_1$ such that the $B_i$ terms can cancel each other for a given initial value of $Q$. For smaller values of $Q$ the $B_i$ terms again will add up to something due to different powers of $Q$. But then the $f_1$ has already been created, and it will remain non-zero all the way to the critical point.

As just noted the $d$ diverges close to $w=0$ with $f_1$ given by expression (\ref{132}). Apart from this uncertainty there is also the location of the end point value of $z$ where $u=m^2=0$. This will change a bit along with the perturbations. Thus the expansions are not properly valid close to $w=0$. However, we will assume that this uncertainty is confined to this region. Also this uncertainty vanishes for small $b_1\rightarrow 0$. But this finite $b_1$ problem is not investigated further in this work. The situation is similar in Sec.~\ref{sec5} for deviations from the critical temperature.

\section{Deviation from critical temperature} 
\label{sec5}

With small deviation $t=\beta-\beta_c$ from inverse critical temperature the perturbation (\ref{115}) will split in another two terms
\begin{equation}
u_1\rightarrow u_1+u_{1t}+u_2+u_{2t}.
\label{134}
\end{equation}
We will assume supercritical temperatures $t<0$ since the HRT equation changes character at phase transitions although critical indices are expected to follow scaling. For the new terms there will be new equations. To be able to cope with the boundary condition (\ref{2}) the  $u_{2t}$ must scale like
\begin{equation}
u_{2t}=tf_{2t}(z)
\label{135}
\end{equation}
with $b'$ linear in $t$, consistent with result (\ref{8}) when  $L$ is given by (\ref{4}).

In general the new term $u_{1t}$ also will be needed. Like the $d$ term of Eq.~(\ref{119}) there will be a $d_t$ term in the equation for $f_{2t}$. It should be linear in $t$ to fit into the scaling form (\ref{135}). This is obtained with products of $u_1$, $u_{1t}$, and their derivatives. Thus like expression (\ref{114}) we get
\begin{eqnarray}
\nonumber
d_t&=&\frac{u_1}{u_0}\frac{u_{1tyy}}{u_{0yy}}+3\left(\frac{u_{1y}}{u_{0y}}\frac{u_{1ty}}{u_{0y}}\right)-2\frac{u_1}{u_0}\frac{u_{1ty}}{u_{0y}}-2\frac{u_{1y}}{u_{0y}}\frac{u_{1tyy}}{u_{0yy}}\\
&+&\frac{u_{1t}}{u_0}\frac{u_{1yy}}{u_{0yy}}+3\left(\frac{u_{1ty}}{u_{0y}}\frac{u_{1y}}{u_{0y}}\right)-2\frac{u_{1t}}{u_0}\frac{u_{1y}}{u_{0y}}-2\frac{u_{1ty}}{u_{0y}}\frac{u_{1yy}}{u_{0yy}}.
\label{136}
\end{eqnarray}
To make this fit into the scaling form of $u_{2t}$ along with form (\ref{116}) for $u_1$ it is required that
\begin{equation}
u_{1t}=\frac{t}{Q^{1/2}}f_{1t}(z).
\label{137}
\end{equation}
With this the equations for the temperature dependence will be similar to Eqs.~(\ref{118}), (\ref{119}), and (\ref{121})
\begin{equation}
-\frac{1}{4}f_{1t}-zf_{1t}^\prime +2F_0\left(\frac{f_{1t}}{f_0}+\frac{f_{1t}^{\prime\prime}}{f_0^{\prime\prime}}-2\frac{f_{1t}^\prime}{f_0^\prime}\right)=0
\label{138}
\end{equation}
\begin{equation}
-zf_{2t}^\prime +2F_0\left(\frac{f_{t2}}{f_0}+\frac{f_{2t}^{\prime\prime}}{f_0^{\prime\prime}}-2\frac{f_{2t}^\prime}{f_0^\prime}+\frac{d_tQ}{t}\right)=0
\label{139}
\end{equation}
for $z<1$, and for $z>1$ they are simply
\begin{equation}
u_{itQ}=0 \quad (i=0,1,2).
\label{140}
\end{equation}

Again like Eq.~(\ref{128}) the obvious solutions for $z>1$ are
\begin{equation}
f_{1t}\propto \frac{1}{z^{1/4}}, \quad f_{2t} \propto \mbox{const.}
\label{141}
\end{equation}
Further with boundary condition (\ref{123}) 
at $z=1$ the solutions for $z<1$ will be like (\ref{132}) and (\ref{133})
\begin{equation}
f_{it}=A_{it}(1+a_i w)+ B_{it}\sqrt{w}, \quad (i=1,2).
\label{142}
\end{equation} 
Again in general $B_{2t}\neq 0$ Thus to fulfill the boundary condition (2) the $B_{2t}$  is to be compensated by $B_{1t}$ for a given starting value $Q$. Then as discussed below Eq.~(\ref{133}) the $f_{1t}$ will be created.

The $u_{1t}$ is the leading scaling function for the temperature dependence from which a second critical index follows and then by scaling the remaining indices. Thus from the $u_0$ of Eq.~(\ref{116}) we have $m^2 \sim Q$ and from (\ref{137}) follows $m^2\sim t/Q^{1/2}$ by which
\begin{equation}
m\sim |t|^{\beta_c}, \quad \beta_c=\frac{1}{3}
\label{143}
\end{equation}
where here $\beta_c$ means the critical index for the curve of coexistence. With this the HRT critical indices in standard notation are
\begin{equation}
\delta=5, \quad \beta_c=\frac{1}{3}, \quad \gamma=\frac{4}{3}, \quad  \alpha=0, \quad \eta=0, \quad \nu= \frac{2}{3}.
\label{144}
\end{equation}
This is expected to hold for subcritical indices too by which they will be $\gamma'=\gamma$, $\alpha'=\alpha$, and $\nu'=\nu$.

The equation of state in the critical region follows from the scaling terms with the subleading contributions considered above. In the limit $Q\rightarrow 0$, i.e.~$z\rightarrow \infty$ they follow from the simple solutions (\ref{128}) and (\ref{141}). So one finds
\begin{eqnarray}
\nonumber
m^2&=&u=u_0+u_1+u_2+u_{1t}+u_{2t}\\
m^2&=&ay^{1/2}+a_1y^{3/4}+by+c\frac{t}{y^{1/4}}+ dt
\label{145}
\end{eqnarray}
where $a$, $a_1$, $b$, $c$, and $d$ are coefficients (in the present case $L=1$ ($z<1$) the $a=6$). This is also the result (plus a couple of higher order terms added) found in Ref.~\cite{hoye11} by accurate evaluations of solutions of the partial differential equation (\ref{1}) for various cutoff functions $L$.

\begin{figure}[h]
  \includegraphics[width=10cm,clip]{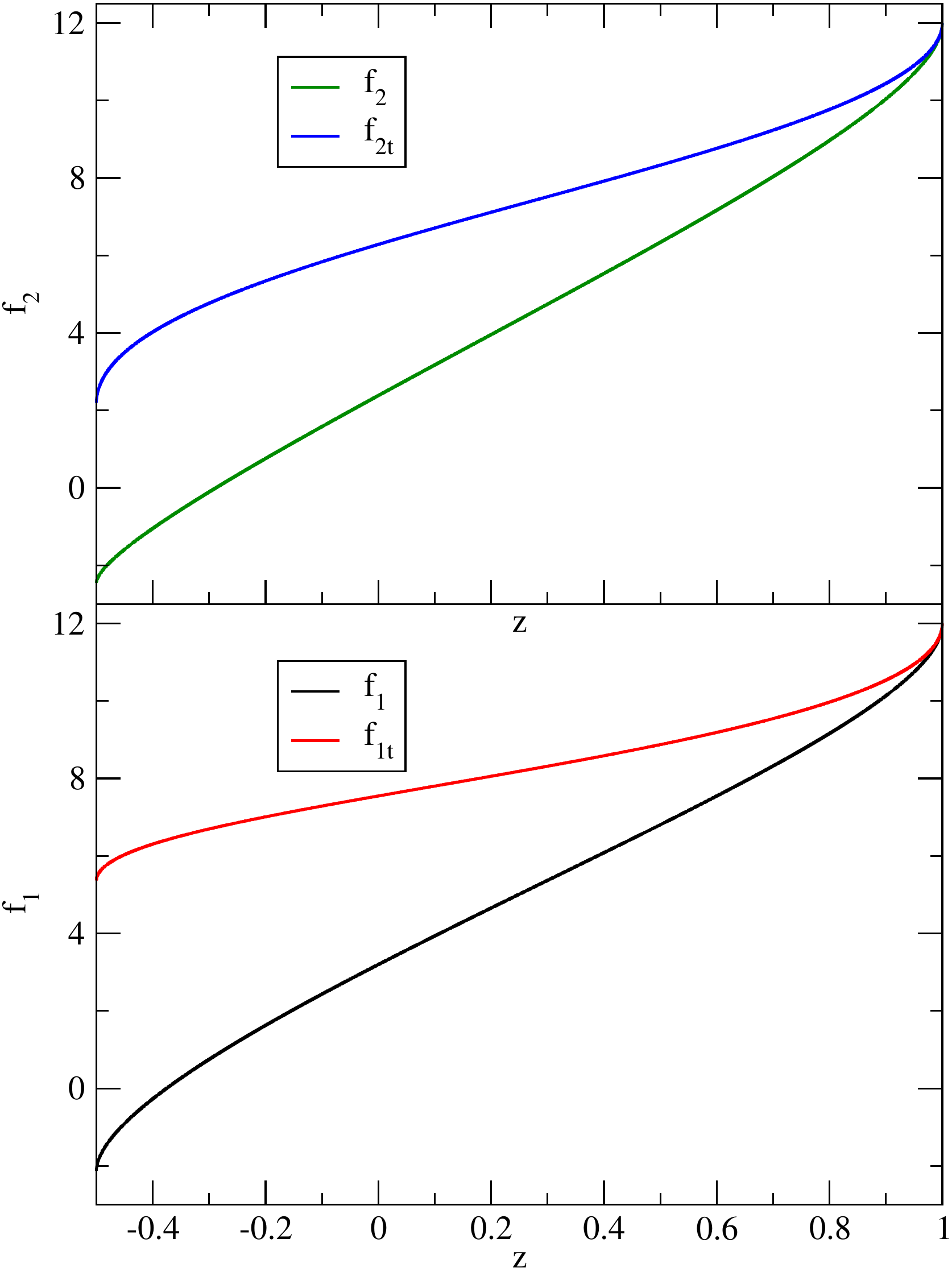}

  \caption{Solutions of Eqs.~(\ref{A9})-(\ref{A12}) including the
    singular part (\ref{A6}), obtained  for parameters $b_1=b_{1t}=1$ for the
functions $f_{1}$ and $f_{1t}$, and then with $b_1=b_{1t}=0.01$ and
$b_2=b_{2t}=1$ for the functions $f_2$ and $f_{2t}$ \label{figfs}.}
\end{figure}

\section{Numerical results} 
\label{sec6}
Our numerical problem thus reduces to solving the differential
equations (\ref{A9})-(\ref{A12}) in the appendix, which correspond to
the regular part of the functions $f_1$, $f_2$, $f_{1t}$, and $f_{2t}$
above. In Appendix.~\ref{secA} the parts singular at $z=1$ have been 
separated out and found as analytic expressions. Now, this set of non-linear second order differential
equations can be cast in the same simple form
\begin{equation}
  \eta(z)f(z)-\delta(z)f'(z)+\alpha(z)(1-z)f''(z)+\gamma(z,f,f',f'') =
  0
  \label{diffeq}
\end{equation}
where, obviously $f$ and its derivatives, stand for  $f_{1r}$,
$f_{2r}$, $f_{1tr}$, and $f_{2tr}$, and the coefficients of the
differential equations are, for $f_{1r}$,

\begin{equation}
  \eta = \frac{3}{4}+\frac{2}{\sqrt{24(1-z)}}; \quad
  \delta=z+\frac{6-\sqrt{24(1-z)}}{3}; \quad \gamma = H_{1s}(z)
  \label{coef1}
\end{equation}
with $H_{1s}$ given by Eq.(\ref{A8}) in the appendix. Similarly, for
$f_{2r}$, $\delta$ stays the same, but
\begin{equation}
   \eta = 1+\frac{2}{\sqrt{24(1-z)}}; \quad \gamma =
   H_{2s}(z)+\frac{2f_0}{\sqrt{24(1-z)}}\left(\frac{d(z,f_1,f_1'.f_1'')}{Q}\right)
   \label{coef2}
\end{equation}
with $H_{2s}$ given by Eq.(\ref{A8}) below and $(d/Q)$ by
Eq.(\ref{114}). In the case of $f_{1tr}$, the coefficients are
identical to those in (\ref{coef1}), but with
\begin{equation}
  \eta = -\frac{1}{4}+\frac{2}{\sqrt{24(1-z)}}; \quad \gamma(z)= H_{1ts}
\end{equation}
with $H_{1ts}$ given by Eq.(\ref{A8}) below. Finally, for $f_{2tr}$, 
\begin{equation}
   \eta = 1+\frac{2}{\sqrt{24(1-z)}}; \quad \gamma =
   H_{2ts}(z)+\frac{2f_0}{\sqrt{24(1-z)}}\left(\frac{d_t(z,f_1,f_{1t})}{Qt}\right)
   \label{coef2s}
\end{equation}
where $H_{2ts}$ given by Eq.(\ref{A8}) below, and $d_t/Qt$ by
Eq.(\ref{136}) where also first and second derivatives of both $f_1$ and $f_{1t}$ are present. The corresponding boundary conditions are
\begin{equation}
  f_{1r}(z=1)=f_{2r}(z=1)=f_{1tr}(z=1)=f_{2tr}(z=1)=0
\label{bca}
\end{equation}
and for the first derivatives we use
\begin{eqnarray}
  \nonumber
  f_{1r}'(z=1) &=& 3b_1; \quad f_{1tr}'(z=1)=-b_{1t}\\
  f_{2r}'(z=1) &=& 4b_2+2b_1^2; \quad f_{2tr}'(z=1) = 0.
  \label{bc}
\end{eqnarray}
where finite second derivatives at boundary are assumed. Also the $\gamma(z=1)$ is neglected, with its
 $1/N$ contribution vanishing. Any inaccuracy here will rapidly disappear 
for decreasing $z$.

\begin{figure}[h]
  \includegraphics[width=10cm,clip]{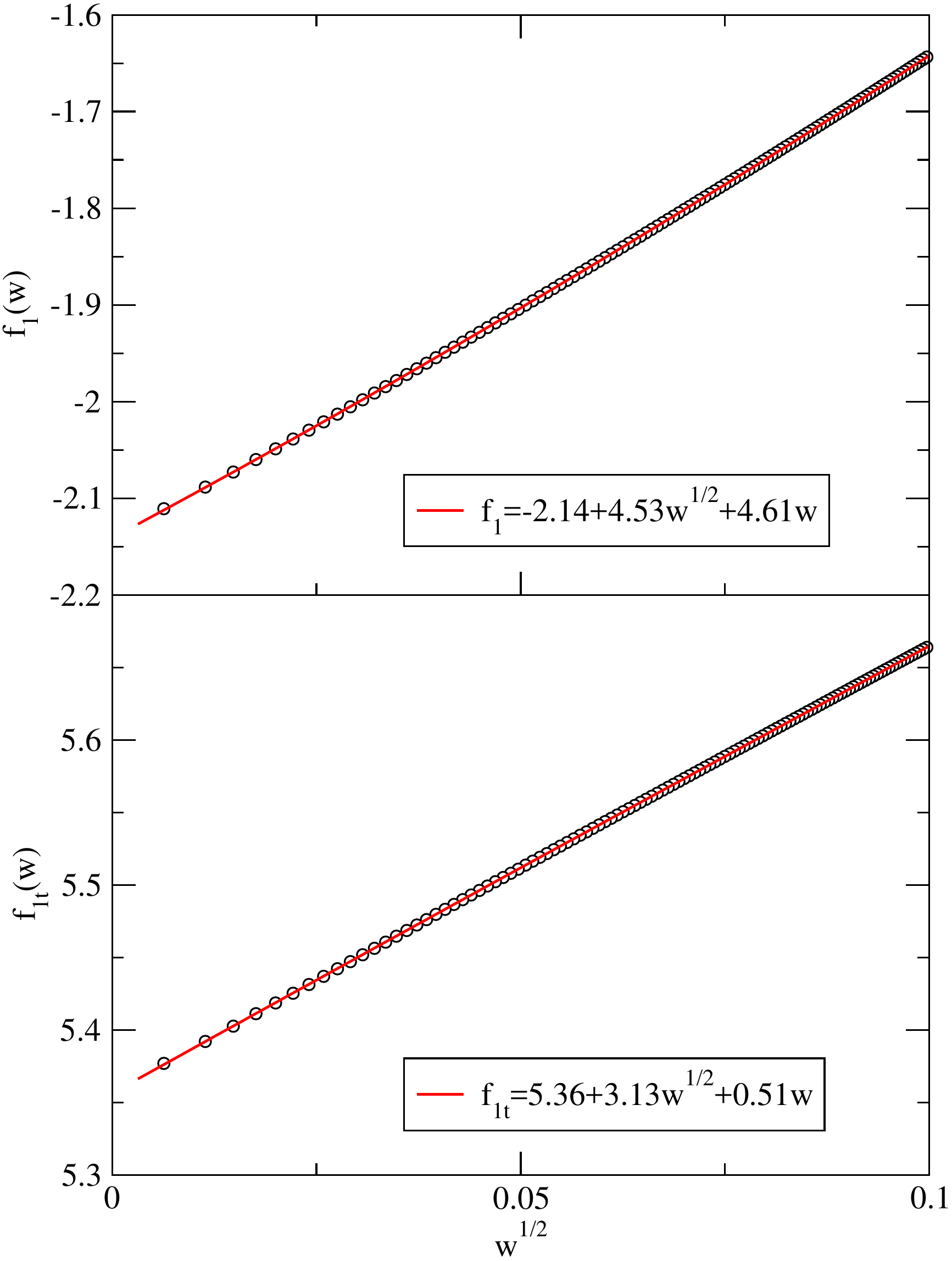}

  \caption{Solutions of Eqs.~(\ref{A9})  and (\ref{A11}) including the
    singular part (\ref{A6}), obtained  for parameters $b_1=b_{1t}=1$ for the
functions $f_{1}$ and $f_{1t}$, when approaching the singularity at
$z=-1/2$ ($w=0$). \label{f1lim}.}
\end{figure}

Now, under these conditions, the differential equations can be solved
sequentially by a simple transformation of the second order
differential equation (\ref{diffeq}) into a system of first order equations,
\begin{eqnarray}
  f'(z) & = & \frac{v(z)}{1-z} \nonumber\\
  v'(z) & = & \left(\delta(z)\frac{v(z)}{1-z}-\eta(z)f(z)-\gamma(z,f,f',f'')\right)/\alpha(z)- \frac{v(z)}{1-z}.
  \label{sys}
  \end{eqnarray}
By discretization  with integration step in $z$, $h$, and using
advanced difference formulas for the derivatives, 
\begin{equation}
  f_k'=\frac{1}{h}(f_k-f_{k-1})=\frac{v_k}{1-z_k},
\end{equation}
one gets
\begin{eqnarray}
  v_{k-1}&=&v_k+h\left(f_k'-(\delta_kf_k'-\eta_kf_k-\gamma_k)/\alpha_k\right)\\
  f_{k-1}&=&f_k-hf_k'
\end{eqnarray}
where $f_N=f(z=1)$, and $f_N'=f'(z=1)$ are given by
Eqs.~(\ref{bca}) and (\ref{bc}). For every step, the values of the second derivatives are
corrected using $f_k''=(f_{k+1}'-f_{k-1}')/(2h)$, ($z_k=i*h$, $N*h=1$,
$0<i<N$), with the values at the boundary obtained by quadratic
extrapolation. The singularities at $z=1$ have been removed from the
functions themselves, but they are still present in the second derivatives,
which implies that a sufficiently small integration step must be used
to guarantee stability. On the other hand, an additional square root
singularity occurs in $f_2$ and $f_{2t}$ when approaching
$z\rightarrow -1/2$, stemming from the $d$ and $d_t$ terms. This can
be mitigated using sufficiently small values of the $b_1$. A more
detailed analysis of this singularity will be
of interest for future work.
\begin{figure}[h]
  \includegraphics[width=10cm,clip]{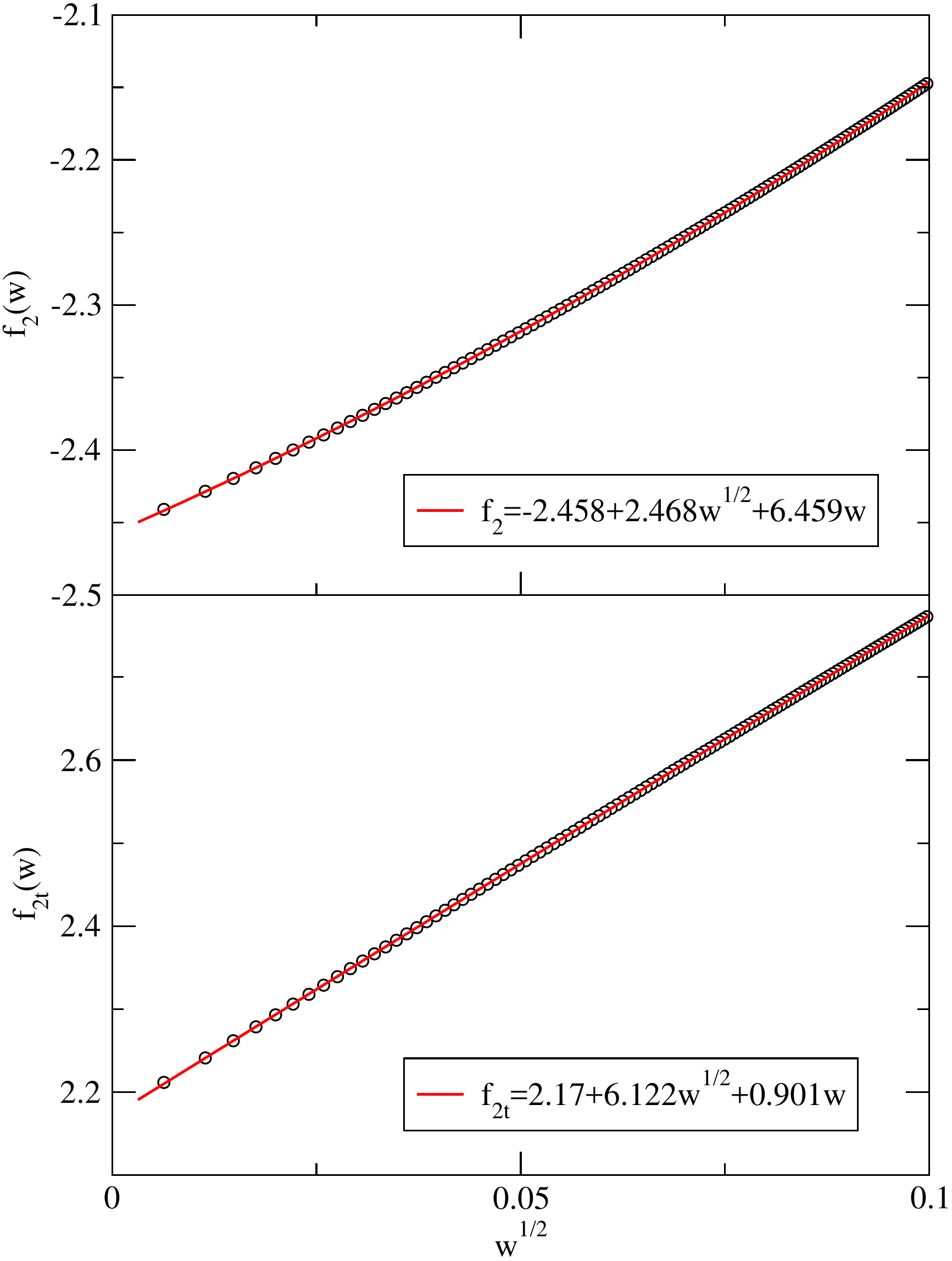}

  \caption{Solutions of Eqs.~(\ref{A10}) and (\ref{A12}) including the
    singular part (\ref{A6}), obtained  for parameters $b_1=b_{1t}=0.01$ and
$b_2=b_{2t}=1$ for the functions $f_2$ and $f_{2t}$ when approaching the singularity at
$z=-1/2$ ($w=0$). \label{f2lim}.}
\end{figure}

Now, using an integration step $h=3\times10^{-6}$, we have solved the
equations (\ref{A9})-(\ref{A12}), first using $b_1=b_{1t}=1$ for the
functions $f_{1r}$ and $f_{1rt}$, and then with $b_1=b_{1t}=0.01$, to tame the
singularity at $z\rightarrow -1/2$, and
$b_2=b_{2t}=1$ for the functions $f_2$ and $f_{2t}$. A general view of the full
solution, including the terms $f_{1s}$, $f_{2s}$, $f_{1ts}$ and
$f_{2ts}$ (see Eq.(\ref{A6}) below), is presented in Fig.~\ref{figfs}.

Now we want to analyze the behavior close to the singularity $z=-1/2$,
where we know from our previous analysis in Secs.~\ref{sec4} and \ref{sec5} that all the functions
basically should follow the form, $f(w) = a+\sqrt{w}+ \ldots$, with
$w=z+1/2$. This form is clearly seen  in Figs.~\ref{f1lim} and \ref{f2lim}.

\begin{figure}[h]
   \includegraphics[width=10cm,clip]{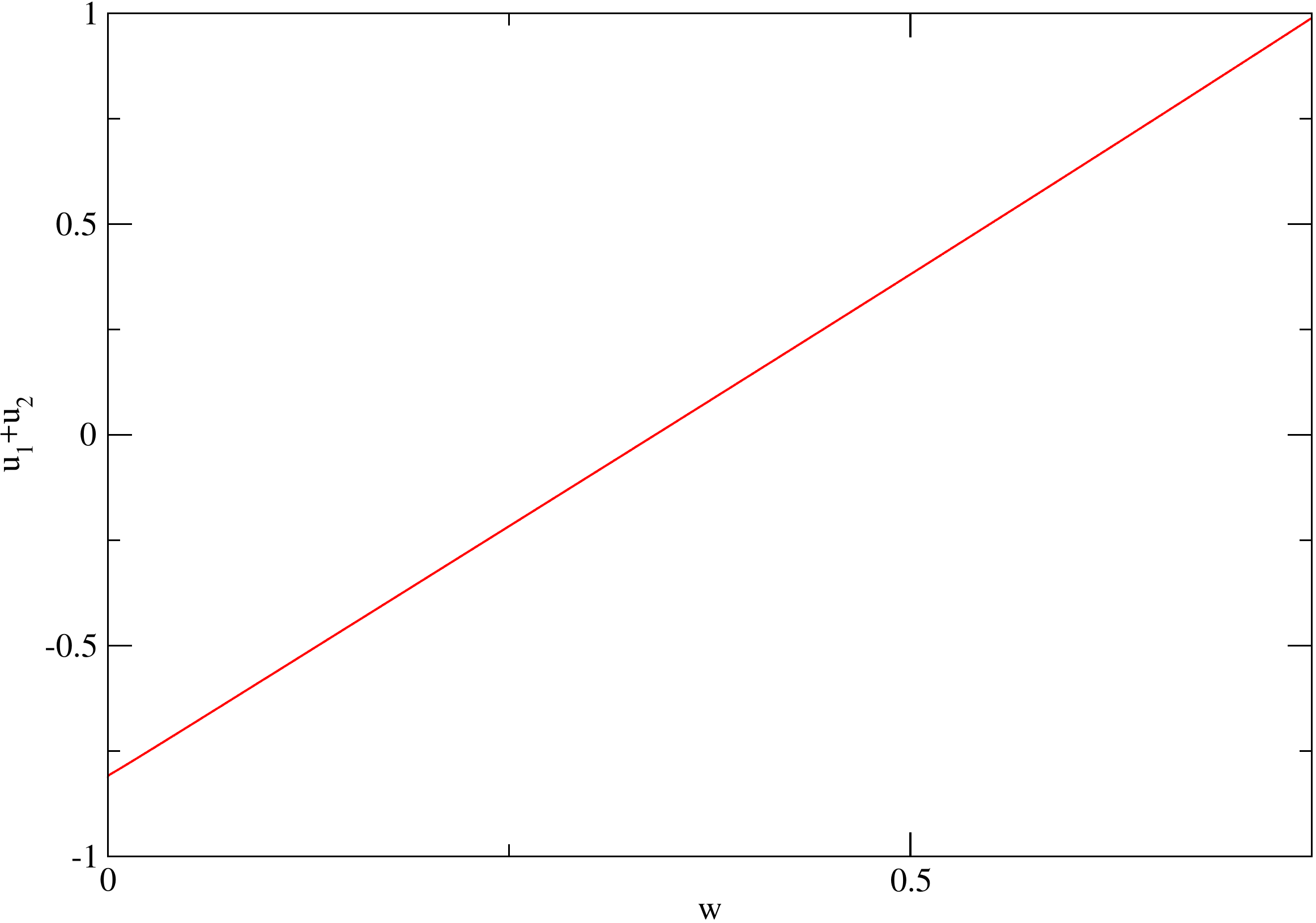}

  \caption{Linear dependence of the sum $u_1+u_2$ when approaching the singularity $z=-1/2$ ($w=0$)
    for the ratio  $b_1/Q^{1/2}=-0.56$.\label{ulim} 
The $u_1$ and $u_2$ are defined in Eq.(\ref{116}), and the $f_2$ used is the one in Fig.~\ref{f2lim}.}
\end{figure}

Finally, taking into account the $Q$ dependence of $u_1$ and $u_2$ in
Eq.(\ref{116}), it is possible to find a ratio of $b_1/Q^{1/2}$, such
that the $w^{1/2}$ dependence of $u_1+u_2$ vanishes and the remaining
term becomes linear in $w$. This is illustrated in Fig.~\ref{ulim},
where we have found that for $b_1/Q^{1/2}=-0.56$ (with $b_2=1$) the square root
singularity disappears, as expected from the analysis of the
equations in Sec.~\ref{sec4}. Further by choosing the starting value of $Q$ small the $b_1$ can also be made small and likewise the $d$ term. 

\section{Summary} 
\label{sec7} 

We have studied the critical properties of the HRT theory by direct
investigation of the HRT partial differential equation where both
analytical and numerical methods have been used. To do so the solution
of a transformed version of the partial differential equation has been
expanded in leading and subleading scaling contributions that fulfill
ordinary differential equations. These contributions are connected
together via simple powers of the cutoff parameter $Q$ of
renormalization. This again leads to simple rational numbers for the
critical indices. 
However, our derivations are not fully valid near $u=m^2=0$ due to
divergence problems discussed at the end of Sec.~\ref{sec4}.
They were not straightened out except in the limit
$b_1\rightarrow 0$. But we do not expect this uncertainty to be
significant for the powers of $Q$ of expansion (\ref{116}) and thus
critical properties. 

The present results confirm those of Ref.~\cite{hoye11} obtained in a
more indirect way by arguments via the somewhat related SCOZA theory
and numerical evaluation of the HRT equation. In a more recent work we
also investigated the critical properties of the HRT of spins in 3
dimensions with spin dimensionality $D$ \cite{lomba14}. There we found
that the critical indices were independent of $D$ for $D$ finite. 
This contrasts earlier studies by renormalization \cite{fisher72} 
and numerical studies of the HRT equation \cite{Parola2012} for this situation where critical indices were found to vary with $D$. But according to Ref.~\cite{lomba14}
the spherical model limit is approached when $D\rightarrow \infty$,
and this has as a consequence that the indices effectively seem to
vary with changing $D$, although they do not. 

We expect that our
present more direct analysis may be applied to the latter situation
too.

\begin{appendix}
\section{Extraction of singular behavior} 
\label{secA}

In our case with $L$ given by Eq.~(\ref{110}) the solutions of equations become singular with diverging derivatives as $z\rightarrow1$, i.e.~$u_y\rightarrow\infty$ with condition (\ref{123}). However, it is possible to extract this behavior with an analytic expression. This will simplify the remaining numerical problem. To do so one may consider the various equations for the subleading contributions. But it turns out that one can just as well solve the non-linear Eq.~(\ref{3}) itself near $z=1$. The solution happens to be a modified version of $u_0$; so we can assume
\begin{equation}
u=(A-BN)Q, \quad  N=\frac{1}{Q}\sqrt{24(Q^2-y)}=\sqrt{24(1-z)}.
\label{A1}
\end{equation}
The coefficients $A$ and $B$ can depend upon Q, and the dominating terms near $z=1$ are
\begin{equation}
u=AQ,\quad u_Q=-B\frac{24Q}{N},\quad u_y=B\frac{12}{N}, \quad u_{yy}=B\frac{144}{N^3}.
\label{A2}
\end{equation}
Inserted in Eq.~(\ref{3}) this gives 
\begin{equation}
\frac{QB}{N}\left(-24+\frac{4A}{B^2}\right)=0, \quad A=6B^2.
\label{A3}
\end{equation}

With expansion in subleading terms of interest one can make the expansion
\begin{equation}
B=1+b_1 Q^{1/2}+b_2 Q+b_{1t} t Q^{-3/2}+b_{2t} t Q^{-1}
\label{A4}
\end{equation}
where $b_1$ etc. are coefficients (or amplitudes), and one finds
\begin{equation}
A=6(1+2b_1Q^{1/2}+(2b_2+b_1^2) Q+2b_{1t} t Q^{-3/2}+(2b_{2t}+2b_1 b_{1t}) t Q^{-1}+\cdots).
\label{A5}
\end{equation}
From this follows the solutions for the f functions
\begin{eqnarray}
\nonumber
f_{1s}=b_1(12-N), \quad f_{2s}=b_2(12-N)+6b_1^2,\\
f_{1ts}=b_{1t}(12-N), \quad f_{2ts}=b_{2t}(12-N)+12b_1 b_{1t}.
\label{A6}
\end{eqnarray}
These singular parts can now be separated out from the $f$ functions, and we are left with the remaining parts $f_r$ that have to be found numerically. The full functions are thus
\begin{equation}
f=f_s+f_r.
\label{A7}
\end{equation}
With solutions (\ref{A6}) for $f_s$ and use of expressions
(\ref{124})-(\ref{126}) for the fixed point solution ($f_0=6-N$ etc.)
the left hand sides of Eqs.~(\ref{118}), (\ref{119}), (\ref{138}), and
(\ref{139}) apart from the $d$ and $d_t$ terms become 
\begin{eqnarray}
\nonumber
H_{1s}&=\frac{1}{4}(36-N)b_1,\quad &H_{2s}=\frac{1}{2}(24-N)b_2+6b_1^2+ \frac{2}{N}6b_1^2,\\
H_{1ts}&=-\frac{3}{4}(4-N)b_{1t},\quad &H_{2ts}=\frac{1}{2}N b_{2t}+ \frac{2}{N}12b_1 b_{1t}.
\label{A8}
\end{eqnarray}
Eqs.~(\ref{118}, (\ref{119}), (\ref{138}), and (\ref{139}) can now be turned into equations for $f_r$, and they become
\begin{equation}
\frac{3}{4}f_{1r}-zf_{1r}' +\frac{2}{N}\left[f_{1r}+f_0\left(\frac{f_{1r}''}{f_0^{\prime\prime}}- 2\frac{f_{1r}'}{f_0'}\right)\right]+ H_{1s}=0
\label{A9}
\end{equation}
\begin{equation}
f_{2r}-zf_{2r}' +\frac{2}{N}\left[f_{2r}+f_0\left(\frac{f_{2r}''}{f_0^{\prime\prime}}- 2\frac{f_{2r}'}{f_0'}+\frac{d}{Q}\right)\right]+ H_{2s}=0
\label{A10}
\end{equation}
\begin{equation}
-\frac{1}{4}f_{1tr}-zf_{1tr}' +\frac{2}{N}\left[f_{1tr}+f_0\left(\frac{f_{1tr}''}{f_0^{\prime\prime}}- 2\frac{f_{1tr}'}{f_0'}\right)\right]+ H_{1ts}=0
\label{A11}
\end{equation}
\begin{equation}
-zf_{2tr}' +\frac{2}{N}\left[f_{2tr}+f_0\left(\frac{f_{2tr}''}{f_0^{\prime\prime}}- 2\frac{f_{2tr}'}{f_0'}+\frac{d_t Q}{t}\right)\right]+ H_{2ts}=0
\label{A12}
\end{equation}
(with $f_0=6-N$, $f_0'=12/N$, and $f''=144/N^3)$. The $d$ and $d_t$ are given by Eqs.~(\ref{114}) and (\ref{136}) respectively. So for $f=f_s$ one finds $f_0 d/Q=-6b_1^2(1+N/2)$ and $f_0 d_tQ=-12 b_1 b_{1t}$ by which the singular $2/N$ terms ($N\rightarrow0$) of (\ref{A8}) are compensated.

With the singular part extracted the boundary condition on the remaining part is simply
\begin{equation}
f_r=0\quad \mbox{for} \quad z=1.
\label{A13}
\end{equation}
for all Eqs.~(\ref{A9})-(\ref{A12}). Since the coefficient of their second derivatives vanishes ($N\rightarrow 0$) as $z\rightarrow 1$ there will be no requirements for specification of their first derivatives; they will follow by solution for decreasing $z$ when starting from $z=1$ when the second derivatives are finite. However, with the numerical method used, initial values (\ref{bc}) are given. As before the solutions for $z>1$ are expressions (\ref{128}) and (\ref{141}) with continuous $f$ functions at $z=1$.

\end{appendix}

\acknowledgments
EL  acknowledges the support from the Direcci\'on
General de Investigaci\'on Cient\'{\i}fica  y T\'ecnica under Grant
No. FIS2013-47350-C5-4-R.

\end{document}